\documentclass{article}
\usepackage{amsmath,amsthm,amsfonts,graphicx}
\usepackage{url,xspace}
\usepackage{rotating}

\newcommand{\GO}[1]{\ensuremath{\mathcal{O}\left(#1\right)}\xspace}

\newcommand{\dqt}{{DQT}\xspace}

\usepackage[ruled,nothing]{algorithm}
\usepackage{algorithmic}

\newcommand{\GF}[1][vide]{\ifthenelse{\equal{#1}{vide}}{}{\ensuremath{\mathtt {GF}(#1)}}}

\newcommand{\Z}{\ensuremath{\mathbb Z}}

\newcommand{\pF}[1][vide]{\ifthenelse{\equal{#1}{vide}}{\Z}{\leavevmode
	\kern.1em\raise.0ex \hbox{\Z}\kern-.1em /\kern-.15em\lower.3ex
         \hbox{#1}\mbox{\Z}}
}

\title{Compressed Modular Matrix Multiplication}

\author{Jean-Guillaume Dumas\footnote{Laboratoire J. Kuntzmann,
    Universit\'e de Grenoble, umr CNRS 5224. BP 53X, 51, rue des
    Math\'ematiques, F38041 Grenoble, France. \{Jean-Guillaume.Dumas,Laurent.Fousse\}@imag.fr} 
\and Laurent Fousse\footnotemark[1] 
\and Bruno Salvy\footnote{Projet ALGO,
INRIA Rocquencourt, 
78153 Le Chesnay.
France. Bruno.Salvy@inria.fr}}

\begin{document}

\maketitle

\begin{abstract} 
  We propose to store several integers modulo a small
  prime into a single machine word. Modular addition is performed by
  addition and possibly subtraction of a word containing several times
  the modulo. Modular Multiplication is not directly accessible but
  modular dot product can be performed by an integer multiplication by
  the reverse integer. Modular multiplication by a word containing a
  single residue is a also possible.
  Therefore matrix multiplication can be
  performed on such a compressed storage. We here give bounds on the
  sizes of primes and matrices for which such a compression is
  possible. We also explicit the details of the required 
  compressed arithmetic routines.
\end{abstract}

\nocite{jgd:2002:fflas,jgd:2004:ffpack,jgd:2004:dotprod,jgd:2007:dqt,jgd:2008:toms}
\nocite{Huang:1979:compress}
\section{Introduction}
 Compression of matrices over fields of characteristic 2 is naturally
 made via the binary representation of machine integers
 \cite{Coppersmith:1993:SLE,Kalto-Lobo}.

 The FFLAS/FFPACK project has demonstrated the need of a
 wrapping of cache-aware routines for efficient small finite field
 linear algebra \cite{jgd:2002:fflas,jgd:2004:ffpack}.
 
 Therefore, 
 a conversion between a modular representation of prime fields of any
 (small) characteristic 
 and e.g. floating points can be performed via the
 homomorphism
 to the integers \cite{jgd:2004:dotprod}. 
 In \cite{jgd:2007:dqt} it is  proposed to
 transform polynomial over a prime field 
 into a $Q$-adic representation where $Q$ is
 an integer than the field characteristic. We call this
 transformation \dqt for Discrete Q-adic Transform.
 With some care, in particular on
 the size of $Q$, it is possible to map the polynomial 
 operations into the floating point arithmetic 
 realization of this $Q$-adic
 representation and convert back using an inverse \dqt.

 Efficient matrix computations over very small finite fields of
 characteristic other than two are required e.g. to study strongly
 regular graphs \cite{May:2007:graphs}, in order to prove/disprove and
 help in the comprehension of the conjectures of \cite{Weng:2007:paley}.

 In this note we propose to use this fast polynomial
 arithmetic within machine words to compute dot products.
 We show in section \ref{sec:dp} how to recover 
 a dot product of size $d+1$ can be recovered from the single 
 coefficient of degree $d$ of a polynomial product.
 Whenever the prime modulus is small enough this enables to compute
 several accumulations of binary products in a single machine
 operation. 
 Then we propose in section \ref{sec:mixed} an alternative
 matrix multiplication 
 using multiplication of a compressed word by a single residue.
 The latter requires also a simultaneous modular reduction, called
 REDQ in \cite{jgd:2007:dqt}.
 In general, the prime field, the size of matrices and the available
 mantissa are given. This gives some constraints on the possible choices
 of $Q$ and $d$. 
 In both cases anyway, we show that these compression techniques 
 represent a speed-up factor of 
 up to the number $d+1$ of residues stored in the compressed format.

\section{Q-adic compression or Dot product via polynomial multiplication}\label{sec:dp}
Suppose that $a(X) = \sum_{i=0}^{d} a_i X^i$ 
and $b(X) = \sum_{i=0}^{d} b_i X^i$
are two polynomials in $\pF[p][X] $. One can perform
the dot product $ \sum_{i=0}^{d} a_i b_{d-i}$ by extracting
the coefficient of degree $d$ of $a(X) b(X)$.

\subsection{Modular dot product via machine word multiplication}

The idea here, as in \cite{jgd:2007:dqt}, is to replace $X$ by an
integer $Q$, usually a power of 2 in order to speed up conversions.
Thus the vectors of residues $a = [a_0 \ldots a_d]$ and
$b=[b_0 \ldots b_d]$ are stored respectively as 
$\bar{b} =  \sum_{i=0}^{d} b_{i} Q^i$ and the {\em reverse} $\bar{a} = \sum_{i=0}^{d} a_{d-i} Q^i$.

This is done e.g. over floating points via the following compressions:
\begin{verbatim}
double& init3( double& r, 
               const double u, const double v, const double w) {
	r=u; r*=_dBase; r+=v; r*=_dBase; return r+=w;
}
\end{verbatim}

\subsection{Gain}
Now for matrix multiplication $A\times B$ 
one wishes to convert a whole row of
the left $m \times k$ matrix $A$ and a whole column of the right
$k\times n$ matrix $B$.
Thus $A$ is transformed into a $m\times \left\lceil \frac{k}{d+1} \right\rceil$
\verb!CompressedRowMatrix!, $CA$ and $B$ is transformed into a
$\left\lceil \frac{k}{d+1} \right\rceil \times n$
\verb!CompressedRowMatrix!, $CB$. 

Therefore the matrix multiply $CA \times CB$ can gain {\em a factor of
  $d+1$} over the multiplication of $A\times B$, for classical
multiplication as shown on the $2\times 2$ example below where the
matrix product {\small $\begin{bmatrix}
  a & b \\
  c & d\\
\end{bmatrix}
\times
\begin{bmatrix}
  e & f\\
  g & h\\
\end{bmatrix}
=
\begin{bmatrix}
 ae+bg  &  af+bh \\
 ce+dg  &  cf+dh \\
\end{bmatrix}$} is performed via integer multiplications. 
The precise gain will be given in table
\ref{tab:gains}.

\[\small
\begin{bmatrix}
  Qa+b \\
  Qc+d\\
\end{bmatrix}
\times
\begin{bmatrix}
  e+Qg & f+Qh\\
\end{bmatrix}
=
\begin{bmatrix}
 * + (ae+bg)Q + *.Q^2  &  * + (af+bh)Q + *.Q^2 \\
 * + (ce+dg)Q + *.Q^2  &  * + (cf+dh)Q + *.Q^2 \\
\end{bmatrix}
\]

The result matrix $C=CA \times CB$ is $m \times n$. Thus in order to
compare similar computations one has either to consider multiplication
of compressed matrices which is then the procedure 
\begin{equation}\label{eq:compC} C=CA\times CB; CC=ReduceAndCompress(C)\end{equation}
or to consider multiplication of normal
matrices via compression and thus the procedure
\begin{equation} CA=CompressRows(A); CB=CompressColumns(B); C=CA\times CB\end{equation}

\subsection{Partial compression}
Note that the last column of $CA$ and the last row of $B$ might not have $d+1$
elements if $\frac{k}{d+1} \notin \Z$. Thus one has to artificially
append some zeroes to the converted values. On $\bar{b}$ this
means just do nothing. On the reversed $\bar{a}$ 
this means multiplying by $Q$ several
times.

\subsection{Delayed reduction and lower bound on $Q$}
For the results to be correct the inner dot product must not exceed
$Q$. With a positive modular representation mod $p$ (i.e. integers
from $0$ to $p-1$), this means that 
$(d+1)(p-1)^2 < Q$. 
Moreover, we would like to use delayed reductions on the intermediate
results and thus accumulate the $\bar{a}\bar{b}$ before any modular
reduction.
It is thus possible to perform matrix multiplications of with common
dimension $k$ as long as:
\begin{equation}\label{eq:lower}\frac{k}{d+1}(d+1)(p-1)^2 = k(p-1)^2<Q. \end{equation}

\subsection{Available mantissa and upper bound on $Q$}
If the product $\bar{a}\bar{b}$ is performed with floating point
arithmetic we just need that the coefficient of degree $d$ remains
fully in the mantissa $\beta$. Write $\bar{a}\bar{b} = c_H Q^d + c_L$, the
latter means that $c_H$, {\em and $c_H$ only}, must remain lower that
$2^\beta$. It could then be exactly recovered by multiplication of $\bar{a}\bar{b}
$ by the correctly precomputed and rounded inverse of $Q^d$ and
floored, as shown e.g. in \cite[Lemma 2]{jgd:2007:dqt}.

With delayed reduction this means that $\sum_{i=0}^d \frac{k}{d+1}
(i+1)(p-1)^2Q^{d-i} < 2^\beta$. We can use equation \ref{eq:lower} in
order to show that 
$\sum_{i=0}^d \frac{k}{d+1} (i+1)(p-1)^2Q^{d-i} \leq Q^{d+1}$. With
this we just have to enforce that 
\begin{equation}\label{eq:upper} Q^{d+1}<2^\beta.\end{equation}

Thus a single reduction has to be made at the end of the dot product
as follows:
\begin{verbatim}
Element& init( Element& rem, const double dp) const {
	double r = dp;
        // Multiply by the inverse of Q^d with correct rounding
        r *= _inverseQto_d; 
        // Now we just need the part less than Q=2^t
        unsigned long rl( static_cast<unsigned long>(r) );
        rl &= _QMINUSONE;
        // And we finally perform a single modular reduction 
        rl %= _modulus;
        return rem = static_cast<Element>(rl);
}
\end{verbatim}

Note that one can avoid the
multiplication by the inverse of $Q$ when $Q=2^t$: 
by adding $Q^{2d+1}$ to the
final result one is guaranteed that the $t(d+1)$ high bits 
represent exactly the $d+1$ high coefficients. On the one hand, 
the floating point multiplication can be replaced by an addition.
On the other hand, this doubles the size of the dot product and thus
reduces by a factor of $\sqrt[d+1]{2}$ the largest possible dot
product size $k$.

\subsection{Results}
\begin{figure}[ht]
\includegraphics[width=\textwidth]{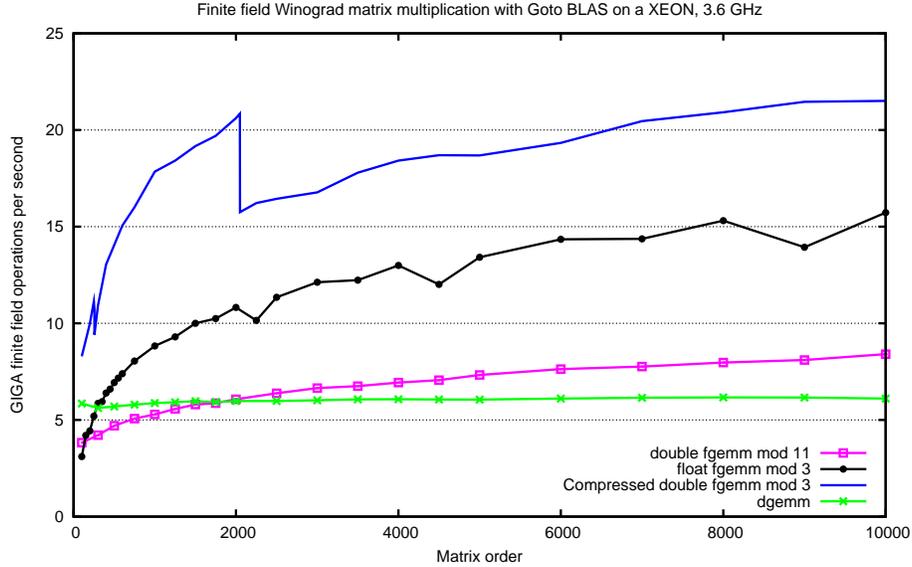}
\caption{Compressed matrices multiplication of equation
  \ref{eq:compC} compared with
dgemm (the floating point double precision 
routine of GotoBlas) and fgemm (the exact
routine of FFLAS) with double or single precision.
}\label{fig:compC}
\end{figure}

One can see on figure \ref{fig:compC} that the compression ($d+1$) is very
useful for small primes since the gain over the double floating point
routine is quite close to $d$. 

Indeed choosing a power of $2$ for $Q$
simplifies and speeds up conversions and thus gives the following
compression factors modulo 3:
\begin{table}[ht]\center\small
\begin{tabular}{|c||c|c|c|c|c|c|c|c|c|}
\hline
Compression & 2 & 3..4 & 5..8 & 8 & 7 & 6 & 5 & 4 & 3\\
\hline
Degree d    & 1 &  &  9 & 7 & 6 & 5 & 4 & 3 & 2\\
\hline
Q-adic      & $2^3$ & $2^4$ & $2^5$ & $2^6$ & $2^7$ & $2^8$ & $2^{10}$ & $2^{13}$ & $2^{17}$\\
\hline
Dimensions  & $2$ & $\leq4 $& $\leq8$ & $\leq16$ & $\leq32$ & $\leq64$ & $\leq256$ & $\leq2048$ & $\leq32768$\\
\hline
\end{tabular}
\caption{Compression factors for different common matrix dimensions
  modulo~3, with $53$ bits of mantissa and $Q$ a power of
  $2$.}\label{tab:comprange} 
\end{table}

Before $n=256$ the compression is at a factor of five and the time to
perform a matrix multiplication is less than a hundredth of a second.
Then from 257 to 2048 one has a factor of 4 and the times are roughly
16 times the time of the four times smaller matrix, as visually shown on
figure \ref{fig:allcomp}, left.
The same is true
afterwards with the respective factor 3 of compression. 

Remark that the curve of fgemm with underlying arithmetic on single
floats oscillates and drops. This is because the matrix begins to be
too large and that modular reductions are required between the
recursive matrix multiplication steps. 
Then the floating point
BLAS\footnote{http://www.tacc.utexas.edu/resources/software/}
routines
are used only when the sub-matrices are small
enough. One can see the subsequent increase in the number of classical
arithmetic steps on the drops around 2048, 4096 and 8192. 

\section{Right Compressed matrix multiplication}\label{sec:mixed}
Another way of performing compressed matrix multiplication is to 
multiply an uncompressed matrix $m \times k$ to the right by a row-compressed
$k \times \frac{n}{d+1}$ matrix.
A dot product with this algorithm will be of the form
$a = [a_0, \ldots, a_n] \times [\sum_{j=0}^d b_{0j}Q^j, \ldots,
\sum_{j=0}^d b_{nj}Q^j]$. 
Therefore, a single entry of the resulting
matrix will be $\sum_{i=0}^k a_i (\sum_{j=0}^d b_{ij}Q^j) =
\sum_{j=0}^d (\sum_{i=0}^ka_i b_{ij}) Q^j$ as shown on the example below. 
\[\small
\begin{bmatrix}
  a & b \\
  c & d\\
\end{bmatrix}
\times
\begin{bmatrix}
  e+Qf\\
  g+Qh\\
\end{bmatrix}
=
\begin{bmatrix}
(ae+bg)+Q(af+bh) \\
(ce+dg)+Q(cf+dh) \\
\end{bmatrix}
\]

Here also $Q$ and $d$ must satisfy equations (\ref{eq:lower}) and
(\ref{eq:upper}).

The major difference is in the reductions. Indeed now one needs to
reduce simultaneously the $d+1$ coefficients of the polynomial in $Q$
in order to get the results. This simultaneous reduction can be made
by the REDQ algorithm of \cite[Algorithm 2]{jgd:2007:dqt}.

Thus the whole right compressed matrix multiplication over two
compressed matrices $CA$ and $CB$, is the following
algorithm as shown also on figure \ref{fig:allcomp}, right:
\begin{equation}\label{eq:LeftComp} A=Uncompress(CA); CC=A\times CB;
  REDQ(CC)\end{equation}

\section{Full compression}
Of course one would like to compress simultaneously two dimensions of
the matrix product. The required dot products can there 
be achieved by polynomial multiplication
with two variables $Q$ and $\Theta$.
Let $d_q$ be the degree in $Q$  and
$d_\theta$ be the degree in $\Theta$.
The dot product is then: 
$a = [\sum_{i=0}^{d_q} a_{i0}, \ldots, \sum_{i=0}^{d_q} a_{in}] \times [\sum_{j=0}^{d_\theta} b_{0j}\Theta^j, \ldots,
\sum_{j=0}^{d_\theta} b_{nj}\Theta^j]$. The latter 
is $\sum_{l=0}^k   (\sum_{i=0}^{d_q} a_{il})(\sum_{j=0}^{d_\theta}
b_{lj})Q^i\Theta^j = \sum_{i=0}^{d_q}\sum_{j=0}^{d_\theta}( \sum_{l=0}^k  a_{il}b_{lj})Q^i\Theta^j $ 
as shown on the example below. 

\[\small
\begin{bmatrix}
  a+Qc & b+Qd \\
\end{bmatrix}
\times
\begin{bmatrix}
  e+\Theta f\\
  g+\Theta h\\
\end{bmatrix}
=
\begin{bmatrix}
(ae+bg)+Q(ce+dg)+\Theta(af+bh)+Q\Theta(cf+dh) \\
\end{bmatrix}
\]

In order to guarantee that all the coefficients can be recovered
independently, $Q$ must still satisfy equation (\ref{eq:lower}) but then
we have this additional equation on $\Theta$:
\begin{equation}\label{eq:qtheta}
Q^{d_q+1} \leq \Theta
\end{equation}
This gives thus upper bounds on $d_q$ and $d_\theta$:
\begin{equation}\label{eq:dq}
Q^{(d_q+1)(d_\theta+1)} < 2^\beta
\end{equation}

\section{Comparison}

\begin{figure}[ht]\hfill
\includegraphics[width=\textwidth*4/11]{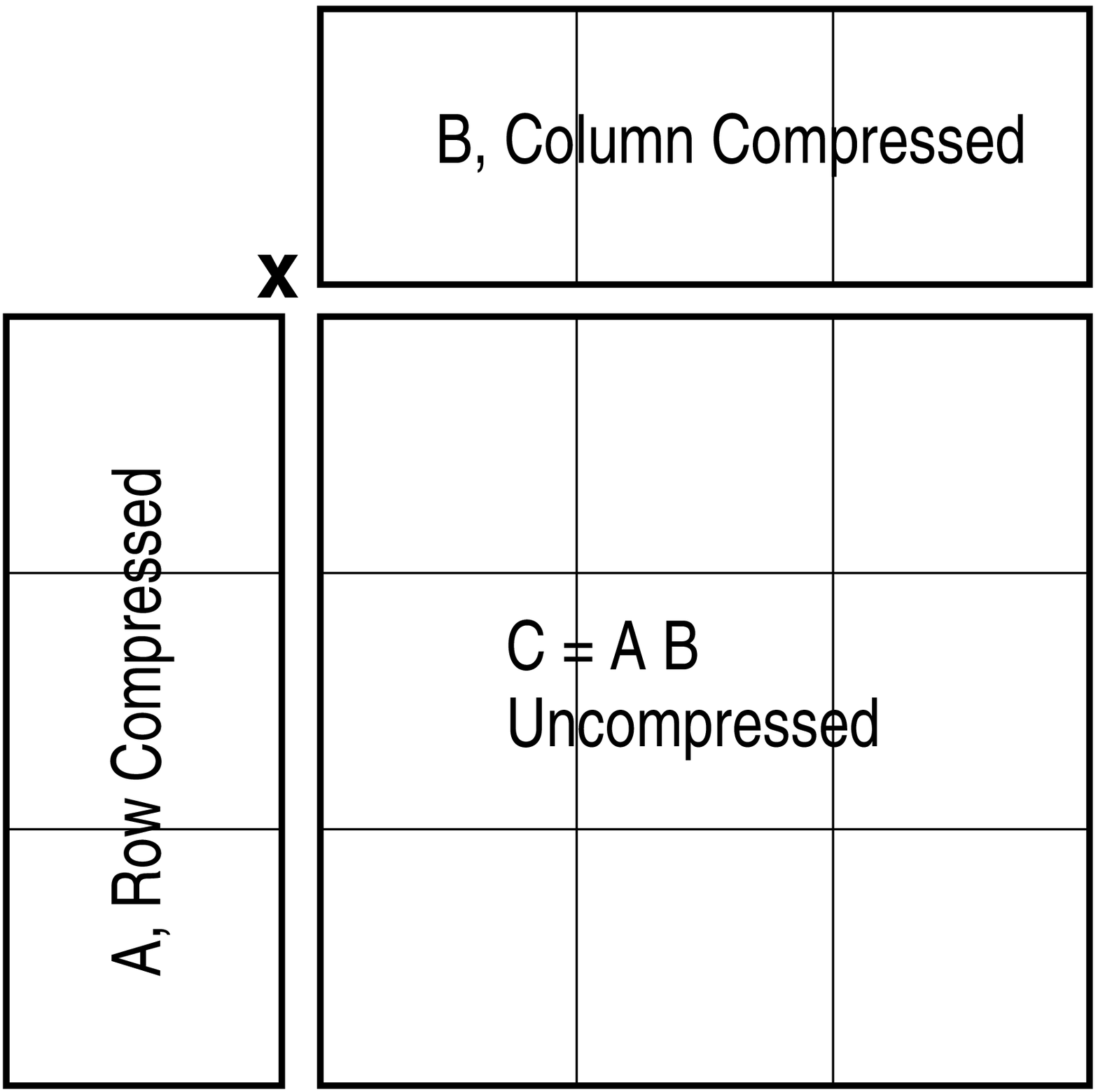}\hfill
\includegraphics[width=\textwidth*4/11]{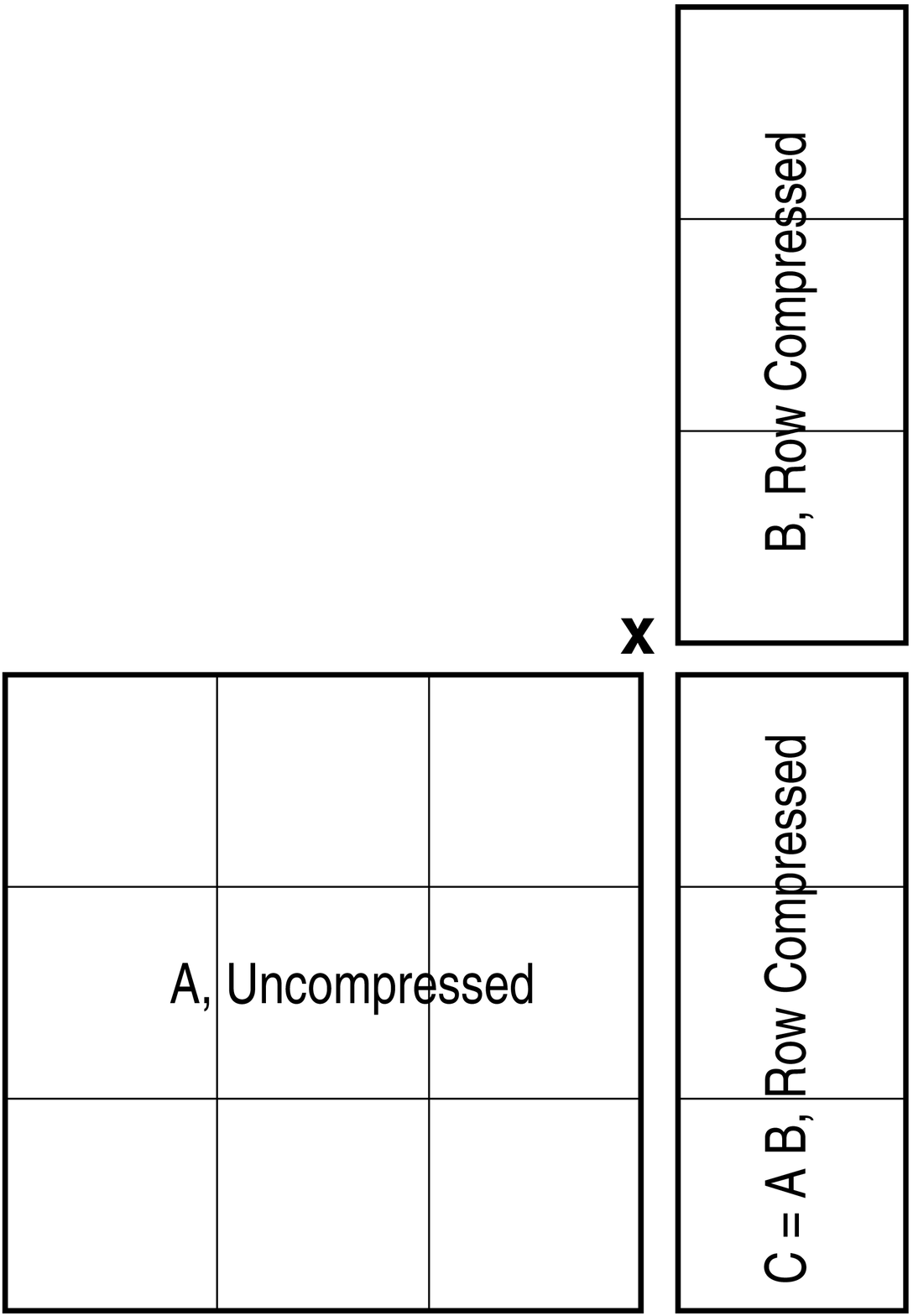}\hfill\ 
\caption{Algorithms of equations (\ref{eq:compC}), left, and
  (\ref{eq:LeftComp}), right.}\label{fig:allcomp}
\end{figure}

\begin{figure}[ht]
\includegraphics[width=\textwidth*6/11]{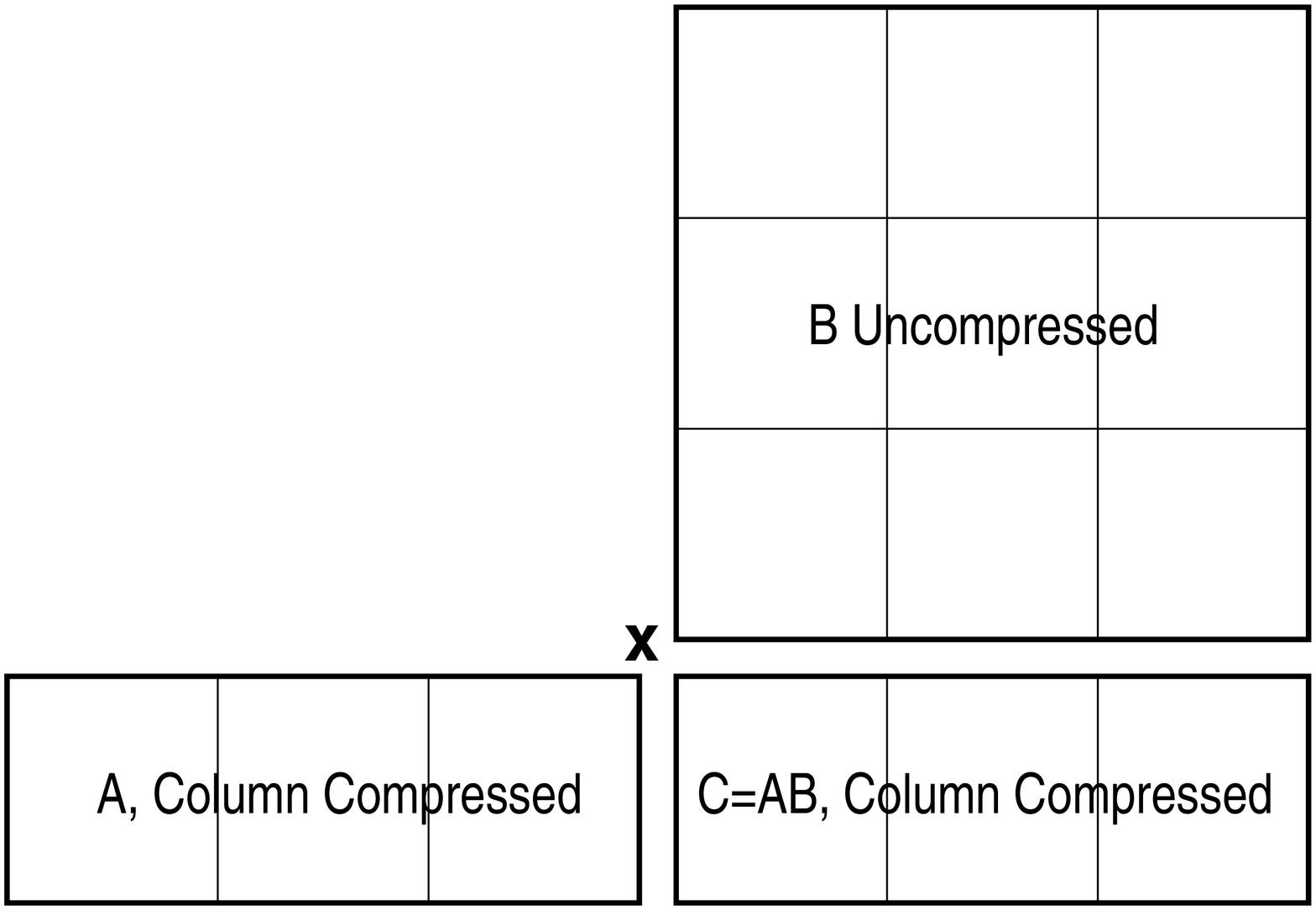}\hfill
\includegraphics[width=\textwidth*4/11]{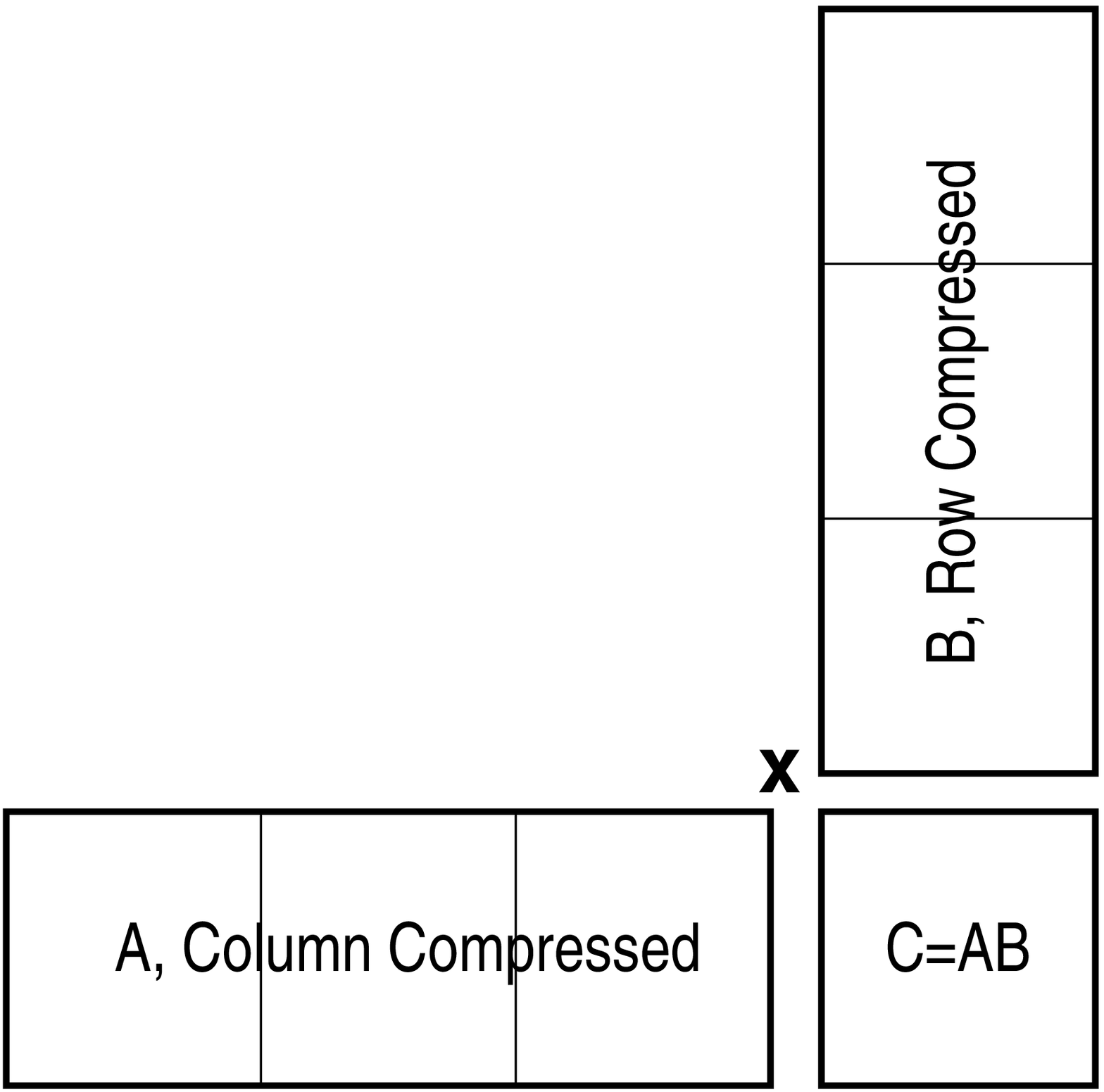}
\caption{Left Compression and Full Compression}\label{fig:fullcomp}
\end{figure}

We summarize the differences of the presented algorithms on figure
\ref{fig:allcomp} and \ref{fig:fullcomp}.

We see on the one hand 
that the first algorithm compresses the common dimension
whereas the Right (or also Left) compressions compress an external matrix
dimension. Thus in the case of rectangular matrices
one can choose between those routines the fastest one. This will be
the routine compressing the largest dimension as shown on table
\ref{tab:gains}. On the other hand the full compression algorithm compresses
both external sizes but, as shown by equation (\ref{eq:dq}) the
available mantissa is is only shared by both compression. Therefore if
we define the {\em compression factor} to be 
\[ e = \left\lfloor \frac{\beta}{\log_2(Q)} \right\rfloor \]
then the degree of compression for the first three algorithms is just
be $d=e-1$ where it becomes $d=\sqrt{e}-1$ for the full compression
with equal degrees for both variables $Q$ and $\Theta$.
For $\omega$ the exponent of matrix multiplication,
the table \ref{tab:gains} 
shows that the gain in terms of arithmetic operations is
$e^{\omega-2}$ for the first three variants and
$e^{\frac{\omega-1}{2}}$ for the full compression.
When $\omega=3$ for classical matrix multiplication the speed-up is
the same. Now, when fast
matrix multiplication is used, as e.g. in \cite[\S
3.2]{jgd:2008:toms},
full compression performs less operations. This is not only of
theoretical interest but also of practical value since the considered
matrices are then less rectangular. This enables more locality for the
matrix computations and usually better performance.

\begin{table}[ht]\small\center
\begin{tabular}{|c||c|c|c|c|}
\hline
Algorithm		& Operations	& Reductions & Conversions\\
\hline
(\ref{eq:compC}) 	& $\GO{ m n
  \left(\frac{k}{e}\right)^{\omega-2}}$ & $m \times n$ $REDC$ &
$\frac{1}{e} mn$ $INIT_e$\\
\hline
(\ref{eq:LeftComp}) 	& $\GO{  m k \left(\frac{n}{e}\right)^{\omega-2}}$ & $m
\times \frac{n}{e}$ $REDQ_e$ & $\frac{1}{e}mn$ $EXTRACT_e$\\
\hline
Left Comp. 	& $\GO{  n k \left(\frac{m}{e}\right)^{\omega-2}}$ & $\frac{m}{e}
\times n$ $REDQ_e$ & $\frac{1}{e}mn$ $EXTRACT_e$\\
\hline
Full Comp. & $\GO{  k \left(\frac{mn}{e}\right)^{\frac{\omega-1}{2}} }$
& $\frac{m}{\sqrt{e}}
\times \frac{n}{\sqrt{e}}$ $REDQ_e$ & $\frac{1}{e}mn$ $INIT_e$ \\
\hline
\end{tabular}
\caption{Number of operations for the different
  algorithms}\label{tab:gains}
\end{table}

The
difference there will mainly be on the number and on the kind of modular 
reductions.
Since the $REDQ_e$ reduction is faster than $e$ classical
reductions, see \cite{jgd:2007:dqt}, and since $INIT_e$ and
$EXTRACT_e$ are roughly the same operations, the best algorithm would
then be one of the Left, Right or Full compression.

For example, with algorithm (\ref{eq:compC}) on matrices of sizes
$10000\times 10000$ it took $92.75$ seconds to perform the matrix
multiplication modulo $3$ and $0.25$ seconds to convert the resulting
$C$ matrix. This is less than $0.3$\%. 
For $250\times 250$ matrices it takes less than $0.0028$ seconds to
perform the multiplication and roughly $0.00008$ seconds for the
conversions. There, the conversions count for $3\%$. 

The full compression algorithm seems the better candidate for the
locality and fast matrix multiplication reasons above ; howbeit the
compression factor is an integer, depending on the flooring of either
$\frac{\beta}{\log_2(Q)}$ or $\sqrt{\frac{\beta}{\log_2(Q)}}$. Thus
there are matrix dimensions for which the compression factor of
e.g. the right compression will be larger than the square of the
compression factor of the full compression. There the right
compression will have some advantage over the full compression.

Further work would thus include implementing the Right or Full
compression and comparing the conversions overhead with that of
algorithm (\ref{eq:compC}).

% \bibliographystyle{plain}
% \bibliography{jgdbibl}

\end{document}